\documentclass[12pt]{article}

\usepackage{graphicx}
\usepackage{slashed}

\newcommand{\bpsi}{\bar{\psi}}

\begin{document}

\title{ \bf    
 Form factors for  pions couplings  to  constituent quarks
under  weak magnetic field
}

 \author{ F\'abio L. Braghin$^1$,  Willian F. de Sousa$^1$,$^2$
\\
%\affiliation{
{\normalsize $^1$ Instituto de F\'\i sica, Federal University of Goias, 
%}
%\\
%{\normalsize 
Av. Esperan\c ca, s/n,
 74690-900, Goi\^ania, GO, Brazil}
\\
{\normalsize $^2$ 
Instituto Federal de Goi\'as,
%}
%\\
%{\normalsize  
R. 75,  n.46, 74055-110, Goi\^ania, GO, Brazil}
}

\maketitle

\begin{abstract}
By considering
a one loop background field method 
for a nonperturbative one gluon exchange
between quarks,
leading electromagnetic  
form factors  of pion couplings  to  constituent quarks are derived
by 
means of   from a large quark  and gluon  effective masses 
expansion. 
With this calculation 
the leading anisotropic  corrections induced by a weak magnetic field 
for the pion- constituent quark 
form factors are obtained.
Besides that, few effective couplings which emerge only due to the weak
magnetic field that break chiral and isospin symmetry 
are also found. 
Numerical estimations  for these  magnetic field corrections
are presented for 
two different  nonperturbative gluon propagators.
All  of these corrections  are ultraviolet finite
and their relative values are found to be of  
order of  $(eB_0/{M^*}^2)^n$ being  $n=1$ 
for the vector 
and axial pion couplings and $n=2$ for the pseudoscalar and scalar ones.
The corresponding  anisotropic corrections to the 
Strong averaged quadratic radii in the 
plane perpendicular to the magnetic field are  also
calculated as functions of the quark effective mass.
\end{abstract}

%\PACS{
%{12.40.Yx}{Hadron mass models and calculations} \and
%{12.38.Lg}{Other nonperturbative calculations}  \and
%{12.39.Fe}{Chiral Lagrangians}  \and
%{14.40.Be}{Light mesons }  
%     }
%}

%{\bf 
%The numerical factors for vector and axial a.q.r. 
%(and also scalar 
%and pseudoscalar ones) 
%turn out to be different due to the explicit isospin-chiral symmetry breaking
%induced by the magnetic field.}

\section{Introduction}
\label{intro}

Although lattice QCD is expected to provide the final
quantitative answers for the description of hadrons in terms of 
the more fundamental degrees of freedom
it is very interesting to develop analytical tools to
describe the gaps between the fundamental level and 
the measured hadron/nuclear properties.
Among these, there are   hadrons  
electromagnetic and strong form factors 
that make possible
a suitable comparison of many important observables 
calculated 
theoretically  from different 
approaches, for example
\cite{beise,drechsel-walcher,maris-craig,yamazaki-etal,constantinou,weise-vogl,hoferichter-etal,ramalho-tsushima-PRD94,eichmann-fischer,exp-bernard+E+meissner,bratt-etal}, 
with experiments,
 for example
\cite{exp-r2A,gaillard-savage,choi-etal-93,bardin-etal-1981,andreev-etal-2007}.
In particular hadron charge distribution,
  spin  structure and 
electroweak interaction properties 
can be understood in terms of electromagnetic and axial form factors.
Due to the enormous difficulties in solving QCD in the low 
energy non perturbative regime,
 effective  models have been developed 
based on general QCD symmetries and properties and  phenomenology,
in particular Chiral Symmetry and  its 
Dynamical  Symmetry Breaking  (DChSB).
The constituent quark model (CQM) 
describes many aspects of phenomenology and it is 
   usually  considered to incorporate the pion cloud
% the  ubiquitous pion as a 
%Goldstone boson
 \cite{constituent-1,constituent-2}.
A further  proposal along these lines is the 
  Weinberg's large N$_c$
effective field theory  (EFT) that copes  constituent quark picture
with 
 the  large N$_c$ expansion 
 \cite{weinberg-2010}.
This EFT has been derived in \cite{EPJA-2016,EPJA-2018}  
without and with 
electromagnetic interaction by starting from a quark-quark interaction
due to a dressed one gluon exchange. 
The resulting effective interactions correspond 
to tree level couplings between pions
 and constituent quarks. 
The
 background quarks, dressed by a sort of gluon cloud described by a 
non perturbative gluon propagator, yield  constituent quarks
\cite{FLB-2018a}.
One might expect that a comparison between the 
electromagnetic and strong  constituent quark form factors and baryons
form factors in the vacuum and at finite energy densities 
might shed light on diverse aspects of baryons structure and interactions
as well as it might provide further criteria to understand 
or improve the reliability of the CQM
to describe low and intermediary energies hadrons.
Besides that, one might 
obtain a systematic way of calculating further 
effects due to   finite energy densities
environments.

In the last decade a high interest on the effect of magnetic fields
on hadron properties and dynamics increased due 
to estimates for  intense magnetic fields  expected to 
appear in peripheral heavy ions collisions, supernovae
and in magnetars \cite{review-1,review-2,nuclei-B}.
Large magnetic fields, 
of the order of $ (e B_0) \simeq 10^{17} -10^{19}$G $\simeq  (0.1 - 15)  
m_{\pi}^2$, 
would not
be so  large as compared to an hadron mass scale
such as the nucleon mass, although it
would only appear for a short time interval
in 
non central heavy ions collisions
\cite{p-rhic}.
One cannot expect large  magnetic fields in low/intermediary energies
hadron collisions in which the usual pion 
dynamics is expected to be dominant, however 
 pion  interactions
 with photons and their behavior under weak magnetic field
might eventually   provide observable effects at intermediary energies.
Moreover, it has been envisaged nuclear structure changes
 due to  external relatively weak magnetic fields  \cite{nuclei-B}
and it becomes 
important to understand further the magnetic field effects
on each part of the nucleon potential among
 which those dictated by the pion couplings.
Many  works have shown effects of magnetic field in 
 hadron properties,  
few examples were given in Refs.
\cite{B-1,B-2,B-3,PRD-2016}.
In Refs. \cite{PRD-2018a,PRD-2018b},
Electromagnetic and Strong form factors of  light  
vector and axial mesons coupled to constituent quarks
 were presented and 
anisotropic corrections to their  quadratic radii
 were found
 due to a weak magnetic field.
%{\bf 
The  pion constituent quark Strong form factors 
in the vacuum
 were presented 
in \cite{FLB-2018a}
and,  in the present work,  their 
 leading photon couplings 
and corresponding  weak magnetic field
 corrections  are  presented
by considering 
the same formalism.
Although one might need strong magnetic fields
to describe completely dynamics in magnetars,
the (relatively) weak magnetic field allows an analytical treatment that
makes explicit magnetic field contributions to Strong interactions observables.

The non perturbative one gluon exchange quark-quark interaction is
one of the leading terms of QCD effective action.
With  the minimal coupling to a background electromagnetic field,
it is given by a  generating functional 
that is usually called global color model given by \cite{PRC1,ER-1986,ERV}:
\begin{eqnarray} \label{Seff}  
Z &=& N \int {\cal D}[\bpsi, \psi]
\exp \;  i \int_x  \left[
\bar{\psi} \left( i \slashed{D} 
- m \right) \psi 
- \frac{g^2}{2}\int_y j_{\mu}^b (x) 
{\tilde{R}}^{\mu \nu}_{bc}  (x-y) j_{\nu}^{c} (y) 
+ \bpsi J + J^* \psi \right] ,
\end{eqnarray}
Where  
 $\int_x$ 
stands for 
$\int d^4 x$,
${\cal D}[\bpsi, \psi]$ is the functional measure of integration,
 $J,J^*$
are the quark sources, $g^2$ is the
 quark gluon coupling constant,
$a,b...=1,...(N_c^2-1)$ stands 
for color in the adjoint representation and 
the quark sources are written in the last terms.
Along the work indices $i,j,k=0,...(N_f^2-1)$ will be used  for  
isospin  indices.
The color  quark current is given by
$j^{\mu}_a = \bar{\psi} \lambda_a \gamma^{\mu} \psi$,
and the sums in color, flavor and Dirac indices are implicit.
$ D_{\mu} = \partial_\mu \delta_{ij} - i e Q_{ij} A_{\mu}$
is the covariant quark derivative  with the minimal coupling to photons,
with  the diagonal matrix 
$\hat{Q} =  diag(2/3, -1/3)$.
The quark current masses $m$ will be considered to be the same
for $u, d$  quarks.
The non perturbative  gluon propagator is an external input
and it is written as 
$\tilde{R}^{\mu\nu}_{ab}(x-y)$.
It must be a non perturbative one
by 
incorporating 
to some extent the gluonic  non Abelian character 
and, in particular,
 it will be required that,
with a corrected quark-gluon coupling,
it  has enough strength to yield dynamical chiral symmetry breaking (DChSB),
as it has been found in several   approches.
Few examples in
\cite{maris-craig,SD-rainbow,kondo,cornwall,higa,holdom,wang-etal}.
 In several   gauges 
this kernel 
 can be written in terms of 
transversal and  longitudinal components,
 as:
$\tilde{R}^{\mu\nu}_{ab} (x-y) \equiv \tilde{R}^{\mu\nu}_{ab} = \delta_{ab} \left[
 \left( g^{\mu\nu} - \frac{\partial^\mu \partial^\nu}{\partial^2}
\right)   R_T  (x-y)
+ \frac{\partial^\mu \partial^\nu}{\partial^2} R_L (x-y) \right]$.

The method employed below has been 
described with details in  
Refs. \cite{EPJA-2016,EPJA-2018}
and therefore
  it will be very succintly reminded in the next section.
%\cite{PRD-2018b,EPJA-2018,FLB-2018a}.
The work is organized as follows.
In the next section the determinant of sea quarks
is presented for structureless pion field
  and then it  is expanded for large quark 
 effective mass and small electromagnetic field. The complete momentum dependence 
of the leading electromagnetic  effective couplings 
of pion interactions with constituent quarks are presented
as momentum integrals of components of 
the quark and gluon propagators. 
Next,  a 
 magnetic field, weak   with respect to  the
 quark effective   mass,  is considered.
The emergence of  anisotropic  corrections to 
 pion-constituent quarks
 couplings  are shown. 
In the following section
two non perturbative
 gluon propagators are considered to provide numerical results,
the Tandy-Maris propagator \cite{tandy-maris}
and an effective confining  one  \cite{cornwall},
both of them produce  DChSB.
The corresponding anisotropic corrections  to the 
Strong (axial and pseudoscalar) quadratic radii due to  a
 weak  magnetic field are also presented.
In the final section there is  a Summary.

\section{ Constituent quarks and quark-antiquark
 light mesons}
\label{sec:two-Q}

To make possible a more complete investigation of all 
the flavor channels,
a Fierz transformation for the model (\ref{Seff}) is performed and
only  the 
 color singlet terms are considered, being 
that the color non singlet ones are non leading order
at least due 
to  a factor $1/N_c$.
Chiral structures with combinations of  bilocal  currents are 
 obtained.
The  quark field must be responsible for the formation of mesons and baryons 
and these different possibilities are envisaged by considering the 
Background Field Method (BFM).
Background quark component will be associated
 to   constituent quark ($\psi_1$) and 
the sea quark can be integrated out ($\psi_2$).
For the one loop BFM it is enough to perform
 a shift of each of the quark currents
 obtained from the Fierz transformation.
For each of the isospin and Dirac channels the following shift
of the bilocal quark currents  is perfomed:
$$\bpsi \Gamma^m \psi \to (\bpsi \Gamma^m \psi)_2 
+ (\bpsi \Gamma^m \psi)_1,
$$
where $\Gamma^m$ are the combinations of 
Dirac and isospin matrices.
 The integration of the sea quark is improved with respect to the 
usual one loop BFM by 
 introducing
light quark-antiquark mesons and excitations by means of 
 the auxiliary field method. 
The following bilocal  auxiliary fields, representing
quark-antiquark states,  are introduced \cite{AFM},
However only the lightest (chiral) 
scalar and isotriplet pseudoscalar sector will be kept analogous 
to Ref. \cite{FLB-2018a} and the heavier vector and axial 
ones will be neglected being less relevant at low energies.
The corresponding   unity integral for the scalar and pseudoscalar
 auxiliary 
bilocal fields $S(x,y), P_i(x,y)$ is the following:
\begin{eqnarray}
 1 &=& {\cal N} \int D[S] D[P_i]
\exp \left( - \frac{i \alpha }{2 }  
\int_{x,y}   R (x-y)   \left[ (S(x,y) - g   j^S_{(2)} (x,y))^2 +
(P_i(x,y) -  g    j^{P}_{i,(2)} (x,y))^2 \right] \right)
,
\end{eqnarray}
where $N''$ is a normalization, the subscripts $_{(2)}$  in the quark currents
stand for the corresponding  quark component and 
$R(x-y) = 3 R_T (x-y) + R_L (x-y)$.
Bilocal auxiliary fields for the  different flavors  
 can be expanded in an infinite orthogonal basis with 
all the excitations in the corresponding channel.
For the pseudoscalar isotriplet fields one has:
%\begin{eqnarray} \label{vec-mes}
\begin{eqnarray}
P_i (x,y) = P_i  \left( \frac{x+y}{2}, x-y \right)
= P_i (u ,z) =  
 \sum_k F_{k} (z) P_{i,k} (u),
\end{eqnarray}
%\end{eqnarray}
where $ F_{k}$ are vacuum functions invariant under translation for
 each of the   local field $P^\mu_{i,k} (u)$.
At lower energies, the lowest energy modes, lighest $k=0$
will be kept, i.e. $P_{i,k=0} = \pi_i$ that are the pions.
The form factors  
 reduce to constants in the zero momentum and structureless  limit
$F_{k}(z) = F_{k} (0)$.
This is obtained  by expanding bilocal fields
 in an infinite orthogonal basis for all the excitations of a given channel, for 
example  for the pseudoscalar isotriplet fields one has:
%\begin{eqnarray} \label{vec-mes}
$P_i (x,y) = P_i  \left( \frac{x+y}{2}, x-y \right)
= P_i (u ,z) =  
 \sum_k F_{k} (z) P_{i,k} (u)$,
%\end{eqnarray}
where $F_{k}(z)$ are vacuum functions invariant under translation for
 each of the   local field $P_{i,k} (u)$.
Only the lowest energy modes, lighest $k=0$ 
can be expected to contribute in
the low energy regime.
In the pseudoscalar channel it 
 corresponds  to dimensionless  pion field:  $P_{i,k=0} = \pi_i$,
and their  form factors  
 reduce to constants  
$F_{k}(z) = F_{k} (0)$ 
endowing the fields with their canonical normalization.
The (heavier) vector and axial mesons
with  their couplings 
to the constituent quarks  were considered
in \cite{PRD-2018a,PRD-2018b}
and  they can be neglected in the 
lower energy regime indeed.
With that, by integrating out the sea quarks, 
the background photon couplings to 
light mesons and constituent quarks arise.
The 
auxiliary fields are undetermined and the
corresponding  saddle point equations
can be used for this.
In the mean field they can be written from the conditions:
\begin{eqnarray}
\frac{\partial S^{eff}_{af} }{\partial \phi_\alpha} = 0,
\end{eqnarray}
where $S^{eff}_{af}$ is the effective action obtained with the integration
of the sea quark with the auxiliary fields and $\phi_\alpha$ stands for 
each of the (constant)  auxiliary fields.
These equations for the NJL model and 
for the model (\ref{Seff})
  have been analyzed in many works
in the vacuum or under  finite energy densities.
In the vacuum, the scalar auxiliary field is the only 
one whose gap equation  has a non trivial solution corresponding 
to a scalar quark-antiquark condensate as the order parameter of DChSB.
At  non zero constant  magnetic fields a   contribution to the 
quark effective mass arises  associated to the so called 
magnetic catalysis that 
is well established from NJL-type and other models and  also 
lattice QCD.
The scalar quark-antiquark field does not necessarily correspond to 
a light  meson and
 a chiral rotation can be performed by freezing this degree of freedom.
Pion field will be described by means of 
the collective variables: $U,U^\dagger=e^{  i \vec{\sigma} \cdot \vec{\pi}}, 
e^{ - i \vec{\sigma} \cdot \vec{\pi}}$.

The sea quark  determinant yields the dynamics of the mesons fields with 
their couplings to constituent quarks, and it 
can be written as \cite{EPJA-2016,EPJA-2018}:
\begin{eqnarray} \label{Seff-det}  
S_{det}   &=&  - i \; Tr  \; \ln \; \left\{  
 i S_{c,q}^{-1} (x-y)
 \right\} 
,
\\
\label{Sc}
S_{c,q}^{-1} (x-y) &\equiv&
 S_{0,c}^{-1} (x-y) 
+ \Xi_s  (x-y)
+
\sum_q  a_q \Gamma_q j_q (x,y) ,
\end{eqnarray}
where 
$Tr$ stands for traces of all discrete internal indices 
and integration of  spacetime coordinates
and $\Xi_s (x-y)$ stands for the coupling of  sea quark to 
the pions.
It  can be written:
% for the two pion fields  respectively as:
\begin{eqnarray} 
    \label{Xi-U}
 \Xi_s  (x-y) &=& F ( P_R  U + P_L  U^\dagger) \; 
  \delta (x - y)
,
\end{eqnarray}
where $P_{R/L} = (1 \pm \gamma_5)/2$ 
are the chiral right/left hand projectors
and $F$ the pion field normalization constant.
The quark kernel can be written
in terms of the effective quark mass generated by the 
scalar field gap equation as 
\begin{eqnarray}
 S_{0,c}^{-1} (x-y)  = \left(  i \slashed{D} -  M^*
\right) \delta (x-y).
\end{eqnarray}
In expression (\ref{Sc}) the following quantity with the color
singlet  chiral
constituent quark currents has been defined:
\begin{eqnarray} \label{Rq-j}
\frac{\sum_q  a_q \Gamma_q  j_q (x,y)}{ \alpha g^2}
&=&
2   R (x-y)
 \left[  \bpsi (y) \psi(x)
+ i  \gamma_5 \sigma_i  \bpsi (y) i \gamma_5  \sigma_i \psi (x)
\right]
\nonumber
\\
&-& 
 \bar{R}^{\mu\nu} (x-y) \gamma_\mu  \sigma_i \left[
 \bpsi (y) \gamma_\nu  \sigma_i \psi(x)
+  \gamma_5   \bpsi  (y)
 \gamma_5 \gamma_\nu  \sigma_i \psi (x) \right] ,
\end{eqnarray}
In this expression
$\alpha=2/9$,  for flavor SU(2) and color SU(3),  and 
 combinations of the longitudinal and transversal parts of the 
gluon propagator
were defined as:
\begin{eqnarray}
R {(x-y)}  &=&
3 R_T (x-y) + R_L (x-y),
\\
\bar{R}^{\mu\nu} {(x-y)}  &=&  g^{\mu\nu}
 (R_T (x-y)+R_L (x-y) )+ 2\frac{\partial^{\mu} \partial^{\nu}}{\partial^2} 
(R_T (x-y) - R_L (x-y) ).
\nonumber
\end{eqnarray}
With the background quark currents, different quark-couplings were found
to emerge in the large quark and gluon effective masses expansion
of the above determinant.
The leading term is an  effective 
(Lagrangian) constituent quark mass that  turns out 
to provide a running quark mass with very similar momentum 
structure to the running quark mass from elaborated
SDE at the rainbow ladder level as shown in \cite{FLB-2018a}.
 This effective mass however is not the same as 
the gap quark effective mass and 
it does not depend on 
DChSB,  being seemingly  akin to other 
mechanisms
\cite{proton-mass}.

\section{ Leading Electromagnetic form factors}

The large effective quark   mass 
expansion of the determinant 
 within  the zero order  derivative expansion
is performed in the following.
The leading momentum dependent couplings  with the 
background electromagnetic 
field   are the following:
\begin{eqnarray}   \label{L-Q-pi-U-A} 
{\cal L}^{q-\pi}_{A} &=&
%\left( 
F_{s,\gamma} (K, Q, Q_1,Q_3)    \;
F_{\mu\nu} (Q_{1})  F^{\mu\nu} (Q_{3})
{\pi}_i (q_{a})  \pi_i (q_{b}) 
 \;
 \bpsi (K)  \psi ( K+Q+Q_1+Q_3 )
\nonumber
\\
&+&
F_{ps,\gamma} (K, Q, Q_1,Q_3)
  \;  F_{\mu\nu} (Q_1) F^{\mu\nu} (Q_3)
  \epsilon_{ij3} \pi_i (Q_1)
 \;
 \bpsi (K) \sigma_j i \gamma_5  \psi (K+Q+Q_1+Q_3 )
\nonumber
\\
&+& 
T_{jki}   F_{V,\gamma} (K, Q, Q_1)
  \;
F^{\mu\nu} (Q_1)   
 \pi_j (q_{a}) 
(\partial_\mu \pi_k (q_{b}) )
 \bpsi (K)   \gamma_\mu \sigma^i \psi (K+Q+Q_1)
\nonumber
\\
&+&  
\epsilon_{ij3}
F_{A, \gamma} (K, Q, Q_1)
  \;
 F^{\mu\nu} (Q_1)  \;  
\partial_\mu \pi_i  (Q) \; 
 \bpsi (K)  i\gamma_5  \gamma_\nu \sigma^j \psi (K+Q+Q_1),
\end{eqnarray}
where in the couplings with two pions  $Q = q_{a}+q_{b}$.
The form factors in this expression
 are given by:
\begin{eqnarray}
F_{s,\gamma} (K, Q, Q_1,Q_3) &=&
  \frac{80}{9}  d_1 N_c (\alpha g^2) e^2 
  F 
   F_{4}^t (K, Q, Q_1,Q_3) ,
\\
F_{ps,\gamma}(K, Q, Q_1,Q_3) &=&
 \frac{64}{3}  d_1 N_c (\alpha g^2) e^2
 \; F  F_{4}^t (K, Q, Q_1,Q_3)
 ,
\\ 
 F_{V,\gamma} (K, Q, Q_1) &=&
 \frac{64}{3} d_1 N_c  F  (\alpha g^2) e \;   F_{5}^t (K, Q, Q_1)
,
\\
 F_{A, \gamma} (K, Q, Q_1) &=&
\frac{64}{3} d_1 N_c  F  (\alpha g^2) e \;  F_{5}^t (K, Q, Q_1) .
\end{eqnarray}

In
 this leading order of the determinant expansion, 
there are also unusual or {\it anomalous}  
couplings to the electromagnetic field 
that also  break chiral and isospin symmetries explicitely and that
correspond to 
 sort of mixing couplings induced by the photon.
 The leading ones can be written as:
\begin{eqnarray}  \label{LA-pij}
{\cal L}_{Aj}
&=& - i
  \epsilon_{ij3} F  {F}_{6P}  (K,Q,Q_1)
A_\mu (Q_1) 
(\partial^\mu \pi_i (q_a)) \pi_j (q_b)
\; \bpsi(K)  \psi (Q_T) 
\nonumber
\\
&-& 
2 i \epsilon_{ij3} F  {F}_{6M}  (K,Q,Q_1) 
A_\mu (Q_1) \partial^\mu \pi_i  (Q) 
\; \bpsi(K) 
i\gamma_5 \sigma^j \psi (Q_T) 
\nonumber
\\
&+&   
i J_{ijk}  F
{F}_{7P}  (K,Q,Q_1) 
A_\mu (Q_1) \; \pi_i (q_a) \pi_j (q_b)  \;
\; \bpsi(K) 
i\gamma^\mu \sigma^k \psi (Q_T) 
\nonumber
\\
&+& 
 2 i  \epsilon_{ji3} F
{F}_{7M}  (K,Q,Q_1)  
A_\mu (Q_1) \pi_j (Q)
\; \bpsi(K) 
i \gamma^\mu \gamma_5 \sigma^j \psi (Q_T) ,
\end{eqnarray}
where
%\begin{eqnarray}
$J_{ii3} = J_{3ii} = - J_{i3i} = 1$,
for  $i=1,2$, 
and $J_{ijk} =  \frac{i}{3} \epsilon_{ijk}$
for $ i\neq j \neq k$.
%\label{J-ijk}
%\end{eqnarray}
The form factors were defined in terms of functions $H_{6P,M}$ and $H_{7P,M}$: 
\begin{eqnarray}
 {F}_{6P}  (K,Q,Q_1) &=&
4 d_1 e N_c  K_0  H_{6P}^t  (K,Q,Q_1),
\\
F_{6M}  (K,Q,Q_1) &=&
4 d_1 e N_c  K_0  H_{6M}^t  (K,Q,Q_1),
\\
{F}_{7P}  (K,Q,Q_1) &=&
4 d_1 e N_c  K_0  H_{7P}^t  (K,Q,Q_1),
\\
{F}_{7M}  (K,Q,Q_1) &=&
4 d_1 e N_c  K_0  H_{7M}^t  (K,Q,Q_1).
\end{eqnarray}

%%%%%%%%%%%%%%%%%%%%%%%%%%%%%%%%%

The functions  $F_i^t (K,Q,Q_1)$, $H_i^t (K,Q,Q_j)$
 used above
were defined as:
\begin{eqnarray}
F_3^t (K,Q,Q_1) &=& \frac{1}{2} \left( F_3 (K,Q,Q_1)  + F_3 (k,Q_1,Q)  \right) ,
\nonumber
\\
F_4^t (K,Q,Q_1,Q_3) &=& \frac{1}{2} \left( F_4 (K,Q,Q_1,Q_3)  + \tilde{F}_4 (K,Q_3,Q_1,Q)  \right) ,
\nonumber
\\  \label{F5t}
F_5^t (K,Q,Q_1) &=& \frac{1}{2} \left( F_5 (K,Q,Q_1)  + F_5 (K,Q_1,Q)  \right) ,
\nonumber
\\
{H}_{6P}^t  (K,Q,Q_1) &=&
\frac{1}{2} \left( {F}_6 (K,Q,Q_1) + {F}_6 (K,Q_1,Q) \right)
,
\\
{H}_{6M}^t  (K,Q,Q_1) &=&
\frac{1}{2} \left( {F}_6 (K,Q,Q_1) - {F}_6 (K,Q_1,Q) \right) ,
\\
{H}_{7P}^t  (K,Q,Q_1) &=&
\frac{1}{2} \left( {F}_7 (K,Q,Q_1) + {F}_7 (K,Q_1,Q) \right)
,
\\
{H}_{7M}^t  (K,Q,Q_1) &=&
\frac{1}{2} \left( {F}_7 (K,Q,Q_1) - {F}_7 (K,Q_1,Q) \right) .
 \end{eqnarray}
In these expressions
the loop momentum  integrals of each of the form factors can be 
written in Euclidean momentum space,
 always for incoming quark momentum $K=0$, as:
\begin{eqnarray}
F_3 (0,Q,Q_1) &=&
\int_k  \left[ k \cdot (k+Q_1) - {M^*}^2 \right]
\tilde{S}_0 (k) \tilde{S}_0 (k+Q_1) \tilde{S}_0 (k+Q+Q_1) \bar{\bar{R}}(-k) ,
\\
F_4 (0,Q_1,Q,Q_3) 
&=&
 \int_k  \left[ -  k \cdot (k+ Q_1+Q) + {M^*}^2 \right]
\tilde{S}_0 (k)  \tilde{S}_0 (k+ Q_1 + Q)
 \tilde{S}_0 (k+Q_1) \tilde{S}_0 (k+Q_4) 
R(- k),
\\
\tilde{F}_4 (0,Q,Q_1,Q_3) &=&
\int_k \left[  k^2 + k\cdot Q - {M^*}^2 \right]
 \tilde{S}_0 (k) 
\tilde{S}_0 (k+Q) \tilde{S}_0 (k+Q+Q_1) \tilde{S}_0(k+Q_4) R(-k) , 
\\ \label{F5}
F_5  (0,Q,Q_1) 
&=&
\int_k  M^*  \tilde{S}_0 (k) \tilde{S}_0 (k+Q_1) 
\tilde{S}_0 (k+Q+Q_1)  \bar{\bar{R}} (-k) ,
\\
F_6 (0,Q,Q_1) &=&
\int_k 
\left[ k \cdot (k +  Q + Q_1 ) - {M^*}^2 \right]
 \tilde{S}_0 (k) \tilde{S}_0 (k+Q_1) \tilde{S}_0 (k+ Q+Q_1) {R} (-k),
\\
F_7 (0,Q,Q_1) &=&
\int_k 
\left[ 3 k^2 + k \cdot (4 Q_1 + 2 Q ) 
 +
Q_5
 - {M^*}^2  \right]
 \tilde{S}_0 (k) \tilde{S}_0 (k+Q_1) 
\tilde{S}_0 (k+ Q+Q_1) \bar{R} (-k),
\end{eqnarray}
where 
$Q_4 = Q_1+Q+Q_3$, $Q_5 =  Q_1^2 + Q \cdot Q_1$,
$\int_k = \int \frac{d^4 k}{(2 \pi)^4}$
and $\bar{\bar{R}}(k)= 2 R(k)$.
The following function was used:
$\tilde{S}_0 (k) = \frac{1}{ k^2 + {M^*}^2}$.

The complete momentum structure of form factors
 such as $F_3$ and $F_4$  present a non monotonic behavior 
that is not  observed experimental data and   they 
yield negative averaged quadratic radii \cite{PRD-2018a}.
To mend this behavior 
  the resulting momentum dependence of the quark kernel 
will be truncated by the following approximation:
\begin{eqnarray} 
S_0^{tr} (k) \sim M^* \tilde{S}_0 (k) .
\end{eqnarray}
It yields truncated form factors  with monotonic behavior with 
pion momenta
that make possible to obtain
always positive  quadratic  mean  radii at this level.
This truncation scheme might be  equivalent to consider
a momentum dependent quark effective mass $M^*$.
Because of that, in the present work only the truncated form 
factors will be investigated numerically.

\begin{figure}[ht!]
\centering
\includegraphics[width=100mm]{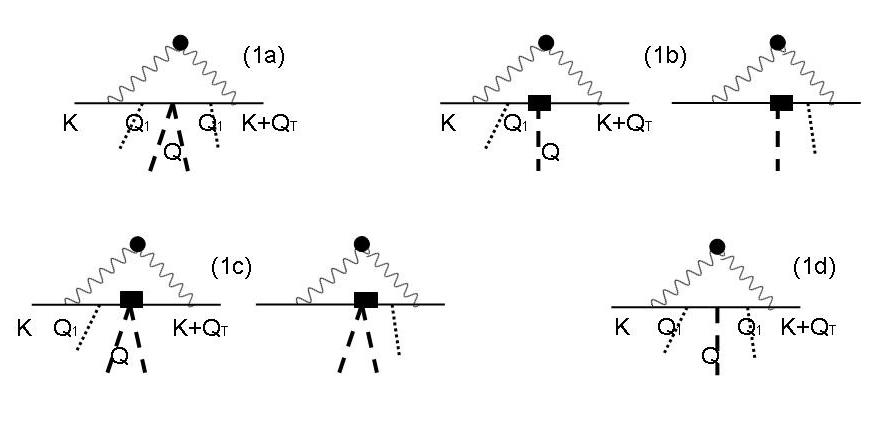}
\caption{ \label{fig:diagrams-2}
\small
Diagrams
(1a,1b,1c,1d)  correspond  to  the couplings of expressions
(\ref{L-Q-pi-U-A},\ref{LA-pij}).
 The wavy line with a full dot is a (dressed) non perturbative gluon propagator,
the solid lines stand for quarks, 
dashed lines for pions
and the dotted line stands 
for the  photon   strength tensor.
A full square in a vertex represent 
momentum dependent pion coupling.
The momenta of each of the particles are indicated by 
$K$ (quarks), $Q$ (pion(s)) and 
$Q_1$ (photons).
}
\end{figure}

The diagrams for  expressions 
%% (\ref{L-Q-pi-W},
(\ref{L-Q-pi-U-A},\ref{LA-pij})
 are presented 
in Fig. (\ref{fig:diagrams-2}).
The incoming constituent quark has momentum  $K$ 
and $K+Q_T$ is  the outgoing constituent quark momentum, where 
$Q_T$ is the total momentum transfered by the pion(s) and photon(s)
to the constituent quark.
$Q$ denotes the total transfered momentum from the pion(s) and 
$Q_1, Q_3$ are each of the  photon momenta.

 \section{ Weak magnetic field}

 It is shown now that 
corrections 
induced by a weak background magnetic field
to those usual form factors
  are obtained
by considering two different effects.
Firstly  the leading correction to the quark kernel and 
also by assuming the weak magnetic field (with respect 
to the constituent quark mass) is strong enough to show up
in the electromagnetic couplings of the previous section.
%For instance $eB_0 \sim 0.1 {M^*}^2 \sim 0.6 m_\pi^2$.
For that, the Landau gauge will be considered in which
  $A^\mu =  - B_0 \; (0,0, x, 0)$.

The 
leading   quark propagator
dependence on the weak magnetic field,  
for equal  up and down quark effective masses $M^*$,
will not be derived in this work and it 
  can be written as  \cite{weak-B,igor1}:
\begin{eqnarray}  \label{quark-prop}
G (k) &=&  S_0 (k) + S_1 (k) (e B_0) = 
\frac{ \slashed{k}+ M^* }{ k^2 - {M^*}^2 } 
+ i \gamma_1 \gamma_2 
 \frac{
 (  \gamma_0 k^0 - \gamma_3 k^3  + M^*  )
 }{
(k^2 - {M^*}^2)^2 }  \hat{Q} (e B_0) .
\end{eqnarray}
 By substituting 
the vacuum quark propagator by
a  $G(k)$ different anisotropic
weak magnetic field-dependent  corrections to pion constituent quark
couplings  appear.
For some of  the leading pion-constituent quark couplings 
found in \cite{FLB-2018a}  however 
the first order correction to the quark propagator $S_1(k)$
does  not contribute in the leading terms, i.e. linear in $(eB_0)$.
In the second order in $(eB_0)^2$
more terms arise but this  will not be considered below.

The correction $S_1(k)$ to the quark propagator
yield the following  anisotropic
  contributions from the leading order determinant 
for the axial constituent quark current
expansion:
\begin{eqnarray} \label{S1-an}
{\cal L}_{S,B} &=&
 G_{A,1}^B \;
 \epsilon^{0 \rho \mu 3}  \;
 \partial_\mu \pi_i \;
 \bpsi  \gamma_5 \gamma_\rho \sigma_i \psi  
\nonumber
\\
&+&
 G_{V,1}^B \;
\epsilon^{ 1 2 \mu \rho}   \; 
\left( \delta_{ij} + \frac{ i }{3} \epsilon_{ijk}
\right) \partial_\mu (\pi_i \pi_j) \;
 \bpsi   \gamma_5 \gamma_\rho \sigma_j \psi  
+ 
G_{V,2}^B \; \partial_0 \pi_i \;  \bpsi   \gamma_z \sigma_j \psi,
,
\end{eqnarray}
where $\epsilon^{\nu \sigma \mu \rho}$ is the Levi Civita tensor, $\partial_0$
stands for $\partial/\partial t$, and 
the effective coupling constants were defined with the trace of 
internal indices as:
\begin{eqnarray} \label{FAUB-exp}
\frac{G_{A,1}^B}{\left( \frac{eB_0}{{M^*}^2} \right)}  =
\frac{2}{3}\frac{ G_{V,1}^B}{\left( \frac{eB_0}{{M^*}^2} \right)}  
= \frac{1}{6} \frac{G_{V,2}^B}{\left(\frac{eB_0}{{M^*}^2}\right)} 
&=&
\frac{8}{3} d_1 N_c  \;  F {M^*}^2 \; (\alpha g^2) \; F_5^t  (0,Q,0) ,
\end{eqnarray}
where $F_5^t(0,Q,0)$ is given above
 in expressions
 (\ref{F5t},\ref{F5}).
 The last  two terms in the second line of (\ref{S1-an}) 
are anomalous since they are not usual Lorentz scalar but pseudoscalar.

Other leading contributions, however,  arise from the zeroth order
quark propagator of expression (\ref{quark-prop}), $S_0(k)$
and a background  photon.
%   $A_\mu =  - B_0 \; (0, x, 0, 0)$.
By considering the pion momentum $Q$, or 
$Q = q_{a}+ q_{b}$ for the two-pion couplings,
the resulting anisotropic 
corrections to pion-constituent quark form factors
are  obtained from expressions
(\ref{L-Q-pi-U-A}),
%% (\ref{L-Q-pi-W},\ref{L-Q-pi-U-A}), 
%% for each of the pion field definitions ($W$ and $U$).
They are given by: 
\begin{eqnarray}
\label{q-pi-U-B}
{\cal L}^{q-\pi}_{B} &=&
  F_s^B(K,Q) 
F {\pi}_i (q_{a})  \pi_i (q_{b}) 
 \;
 \bpsi (K)  \psi (K+Q )
\nonumber
\\
&+&
\epsilon_{ij3}
F_{ps}^{B} (K,Q)
F  \pi_i (Q)
 \;
 \bpsi (K) \sigma_j i \gamma_5  \psi (K+ Q )
\nonumber
\\
&+& 
T_{jki}   
F_{V}^{B} (K,Q)
 \pi_j (q_{a}) 
(\partial_\mu \pi_k (q_{b}) )
 \bpsi (K)   \gamma_\mu \sigma^i \psi (K,Q)
\nonumber
\\
&+& 
  \epsilon_{ij3}\;   
F_{A}^{B}  (K,Q) \;
\partial_\mu \pi_i  (Q) \; 
 \bpsi (K)  i\gamma_5  \gamma_\nu \sigma^j \psi (K,Q)
,
\end{eqnarray}
where 
$ T_{jki} = \delta_{ij}\delta_{3k} - \delta_{j3} \delta_{ik}$
and 
\begin{eqnarray} \label{ff-W-vec-B}
\label{ff-sca-B}
F_{ps}^{B} (K,Q)
= \frac{12}{5}  F_s^B (K,Q)  
&=&   
\left( \frac{e B_0}{{M^*}^2}
\right)^2
 \frac{64}{3}  d_1 N_c {M^*}^4
 (\alpha g^2)   \; F_{4}(K, Q, Q_1=Q_3=0)
,
\\  \label{FAUB} \label{ff-U-vec-B}
F_{A}^{B} (K,Q) 
= \frac{F^B_V}{3} (K,Q)
&=&
\left( \frac{e B_0}{{M^*}^2}
\right)
 \frac{64}{3}
 d_1 N_c  F {M^*}^2  (\alpha g^2) \;  F_{5}(K, Q, Q_1=0)
.
\end{eqnarray}
The slightly more  symmetric  way of defining the magnetic field
$A^\mu=-B_0(0,y,x,0)/2$ garantees the anysotropies to 
be kept in the plane perpendicular to the magnetic field.
The quark effective mass $M^*$ receives corrections 
from the weak  magnetic field in the scalar
gap equation and 
it will not addressed with  details  in the present work.
These expressions provide numerical values
one order of magnitude smaller than
of the original 
pion - constituent quark couplings
 because they have 
multiplicative extra factors $B_0$ or $B_0^2$ that can be factorized
in constants such as
as $eB_0/{M^*}^2$
or $(eB_0)^2/{M^*}^4$ within  the current  large quark effective mass regime.
These make explicit that the $B_0$ induced corrections are considerably smaller 
than the original coupling constants and 
form factors.

The {\it anomalous}
 electromagnetic pion-quark couplings
 from expression (\ref{LA-pij})
generate  magnetic field induced  mixing pion  couplings
breaking explicitely chiral and isospin symmetries.
For pion momentum $Q$, or  $Q=q_a+q_b$ 
in the two pion couplings,
it yields:
\begin{eqnarray}   \label{L-Aj-B}
{\cal L}_{Aj}
&=&
-  
 i  \epsilon_{ij3} {F}_{6PB}  (K,Q)
(\partial^y \pi_i (q_a)) \pi_j (q_b)
\; \bpsi(K)  \psi (K+Q) 
\\
&-& 
2 i \epsilon_{ij3}  {F}_{6MB}  (K,Q) 
\partial^y \pi_i  (Q) 
\; \bpsi(K) 
i\gamma_5 \sigma^j \psi (K+Q) 
\nonumber
\\
&+&   
i J_{ijk} 
{F}_{7PB}  (K,Q)
 \; \pi_i (q_a) \pi_j (q_b)  \;
\; \bpsi(K) 
i\gamma^2 \sigma^k \psi (K+Q) 
\nonumber
\\
&+& 
 2 i  \epsilon_{ji3} 
{F}_{7MB}  (K,Q)  
\pi_j (Q)
\; \bpsi(K) 
i \gamma^2 \gamma_5 \sigma^j \psi (K+Q) .
\nonumber
\end{eqnarray}
The following functions were used in expressions (\ref{L-Aj-B})
written in terms of the momentum 
derivative $\partial_{Q_{1,x}}=\partial/\partial Q^x_1$:
\begin{eqnarray}
F_{6BP} (0,Q) &=& 
\left( \frac{e B_0}{{M^*}^2}
\right)
4 d_1 F N_c   {M^*}^2 K_0
 \left( \partial_{Q_{1,x}}  \; 
  H_{6P}^t (0,Q,Q_1) \right)_{Q_1=0} ,
\\
F_{6BM} (0,Q) &=&  
\left( \frac{e B_0}{{M^*}^2}
\right)
4 d_1  F  N_c  {M^*}^2 K_0 \; 
 \left( 
\partial_{Q_{1,x}}   
 H_{6M}^t (0,Q,Q_1) \right)_{Q_1=0} ,
\\
F_{7BP} (0,Q) &=&
\left( \frac{e B_0}{{M^*}^2}
\right)
4 d_1    F N_c  {M^*}^2 K_0  \; 
\left(  \partial_{Q_{1,x}}  
 H_{7P}^t  (0,Q,Q_1) 
  \right)_{Q_1=0} ,
%\nonumber
\\
F_{7MP} (0,Q) &=& 
\left( \frac{e B_0}{{M^*}^2}
\right)
4 d_1  F  N_c  {M^*}^2 K_0  \;
\left(  
\partial_{Q_{1,x}}   
 H_{7M}^t  (0,Q,Q_1) 
  \right)_{Q_1=0} .
\end{eqnarray}

By comparing expressions (\ref{FAUB-exp}) 
and 
(\ref{FAUB})
the following exact ratios are 
 obtained:
\begin{eqnarray}
\frac{G_{A,1}^B(0,0)}{F_A^{B}(0,0)} = 
2 \frac{G_{V,1}^B(0,0)}{F_V^{B}(0,0)}
= \frac{G_{V,1}^B(0,0)}{G_{V,2}^B(0,0)}
= \frac{1}{8}.
\end{eqnarray}
These  ratios are  
 simple numerical factors 
and the
momentum dependence 
of only one of these form factors, $F_A^{B}(K,Q)$,
 will be explicitely shown below.

\subsection{ Numerical results}

In the following, 
numerical estimations for the 
form factors above
%(\ref{q-pi-W-B},\ref{q-pi-U-B},\ref{L-Aj-B})
 will be shown.
Two gluon propagators will be considered,
 firstly a transversal  one  from Tandy-Maris $D_{I}(k)$
\cite{tandy-maris}
and  the other is an  effective confining 
 longitudinal one  by Cornwall 
$D_{II}(k)$
 \cite{cornwall}.
Both of them 
yield DChSB in the gap equation for the scalar auxiliary field
being that:
\begin{eqnarray} \label{propagators}
g^2 \tilde{R}^{\mu\nu} (k) \equiv  h_a D_a^{\mu\nu} (k) ,
\end{eqnarray}
where $D^{\mu\nu}_a(k)$ ($a=I,II$) 
are  the chosen  gluon propagators whose 
expressions 
 are the following:
\begin{eqnarray}
D_I (k) &=& 
\frac{8  \pi^2}{\omega^4} De^{-k^2/\omega^2}
+ \frac{8 \pi^2 \gamma_m E(k^2)}{ \ln
 \left[ \tau + ( 1 + k^2/\Lambda^2_{QCD} )^2 
\right]}
,
\\
D_{II} (k) &=& 
K_F/(k^2+ M_k^2)^2 ,
\end{eqnarray} 
where
for the first expression 
$\gamma_m=12/(33-2N_f)$, $N_f=4$, $\Lambda_{QCD}=0.234$GeV,
$\tau=e^2-1$, $E(k^2)=[ 1- exp(-k^2/[4m_t^2])/k^2$, $m_t=0.5 GeV$,
$\omega = 0.5$GeV, $D= 0.55^3/\omega$ (GeV$^2$); 
and for the second expression
$K_F = (  2 \pi  M_k / (3 k_e) )^2$
where  $k_e  = 0.15$   and $M_k = 220$MeV. 
In expression (\ref{propagators}) $h_a$ is a constant factor
already considered in previous works \cite{EPJA-2018,PRD-2018b,FLB-2018a}
to fix the quark gluon (running) 
coupling constant such as
to reproduce a particular value for one of the resulting effective 
coupling constant, for example the pion axial coupling constant
$g_A h_a = 1$ or the 
rho  constituent quark coupling constant $g_\rho h_a \simeq 12$.
 In the present work they 
were fixed  with the same convention of 
the pion  constituent quark form factors of Ref. \cite{FLB-2018a}
for each of the gluon propagators, 
 $h_I=0.82$ and $h_{II}=0.3$ .

\begin{figure}[ht!]
\centering
\includegraphics[width=140mm]{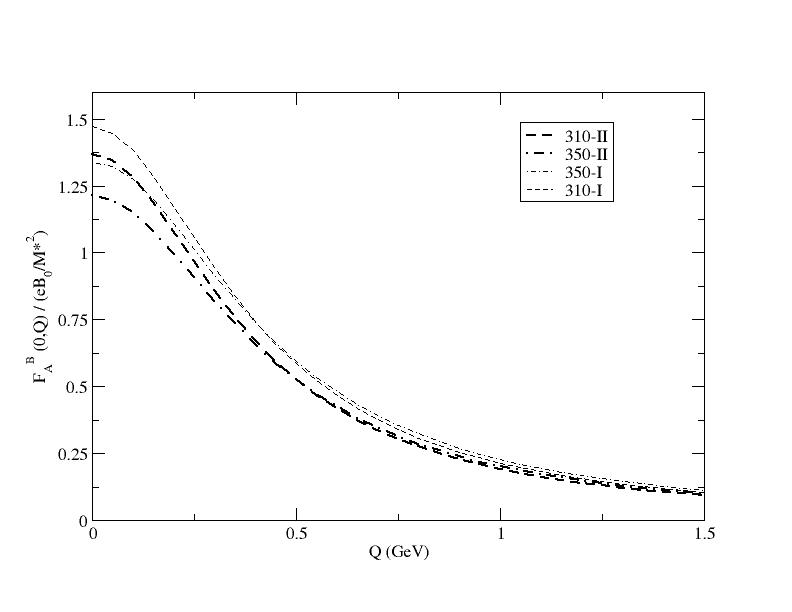}
%% {F-piA-B-U-TM-CO-rat-nov-ha.jpg}
%% ]{F-piA-B-U-TM-CO-rat-new-2.jpg}
%% {F5B-CO-TM.jpg}
\caption{ \label{fig:FPIABUTM}
\small
In this figure the leading magnetic field correction to the axial 
 form factor 
$F_{A}^{B} (0,Q)/(eB_0/{M^*}^2)$ as a function of the pion momentum
for the usual pion field
 is presented 
for the two gluon   propagators $D_I (k)$  (thin lines) and 
 $D_{II}(k)$ (thick lines).
Different values of the  quark effective mass 
are considered
$M^*=350$MeV in dot-dashed lines, $M^*=310$MeV in dashed  lines.
 $h_I=0.82$ and $h_{II}=0.3$
}
\end{figure}

In figure (\ref{fig:FPIABUTM})
 the axial form factor
$F^{B}_{A}(K,Q)/(eB_0/{M^*}^2)$
is shown as function of the pion momentum $Q = |Q|$
for two different quark effective mass $M^*$
This contribution for the form factor  in the figure is divided by 
a factor  $(e B_0)/{M^*}^2$
to make easier the interpretation 
and the comparison of each of the contributions
for any small value of $(e B_0)/{M^*}^2$.
The effective mass $M^*$ is, however, kept constant.
At zero pion momentum $Q=0$ the form factor yields the axial 
pion coupling to constituent quark that is usually considered to be
$g_A=0.8$ or  $1$ \cite{weise-vogl,weinberg-2010}.
 Therefore, without further assumptions about the 
quark-gluon coupling constant,
the weak magnetic field correction to the axial form factor 
is smaller for the gluon propagator $D_{II}(k)$ than
for $D_I(k)$ and this is simply  due to the overal strength of the 
corresponding propagator and 
 quark-gluon coupling constant.

\begin{figure}[ht!]
\centering
\includegraphics[width=140mm]{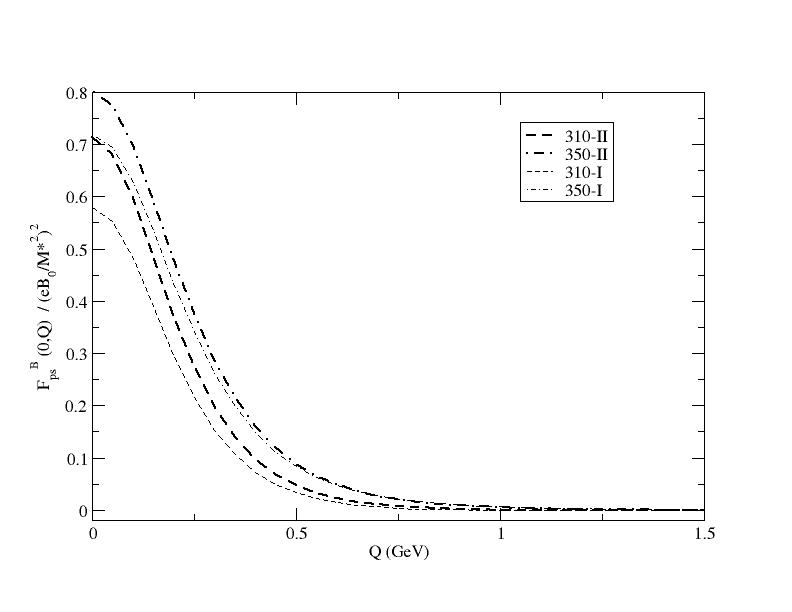}
\caption{ \label{fig:FPSPHBTMCO}
\small
In this figure the leading magnetic field correction to the 
 pseudoscalar form factor 
$F_{ps}^{B} (0,Q)/(eB_0/{M^*}^2)^2$ 
as a function of the pion momentum
for the usual  pion field
 is presented 
for the two gluon   propagators $D_I (k)$  (thin lines) and 
 $D_{II}(k)$ (thick lines).
Different values of the sea quark effective mass 
are considered
$M^*=350$MeV in dot-dashed lines,
 $M^*=310$MeV in dashed  lines.
%$M^*=280$ MeV in solid lines
% $M^*=70$ MeV in dotted lines.
 $h_I=0.82$ and $h_{II}=0.3$
}
\end{figure}

In figure (\ref{fig:FPSPHBTMCO}) the weak magnetic field anisotropic 
correction 
to the pseudoscalar form factor, divided by $(eB_0/{M^*}^2)^2$,  
is presented for the two gluon propagators as a function of the pion momentum
 and for different values of $M^*$.
Although the value of the pseudoscalar coupling constant 
is of the order of 10 times the
value of the axial coupling constant
the weak magnetic field corrections calculated in these figures are 
basically of the same order of 
magnitude.

\begin{figure}[ht!]
\centering
\includegraphics[width=140mm]{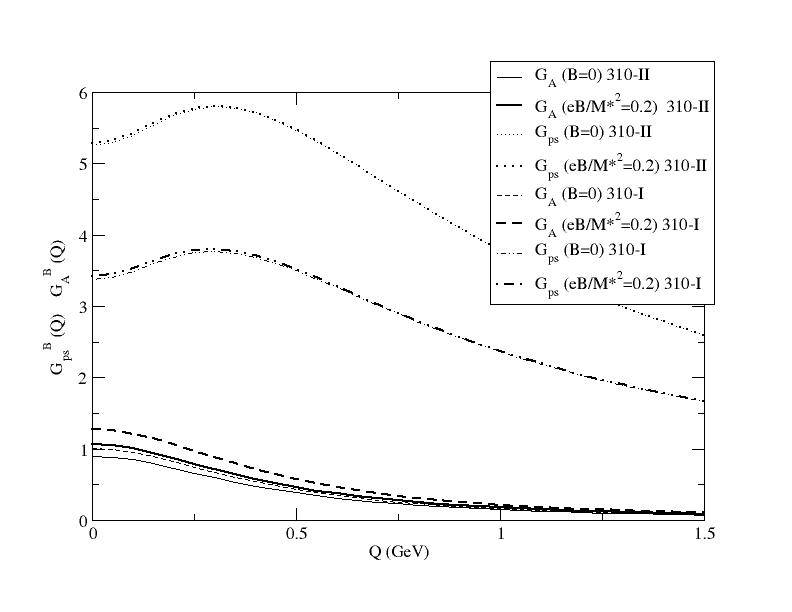}
%% {Gps-GA-0+B-nov-trunc.jpg}
\caption{ \label{fig:GPS-GA}
\small
Pseudoscalar and axial (non truncated) form factors  for
$B_0=0$, from \cite{FLB-2018a}, 
 and for  $(eB_0/{M^*}^2) = 0.2$
with different contributions: from the quark mass dependence on 
the weak magnetic field and from the correction  to the form factor 
from expressions (\ref{GA-ff-B},\ref{gps-Bfield}) 
for  the two 
  gluon propagators $D_I(k)$ 
and $D_{II}(k)$ and
with 
$M^*=310$MeV.
 $h_I=0.82$ and $h_{II}=0.3$.
}
\end{figure}

A complete expression  for  the axial and pseudoscalar
 form factors, including
both their values in the vacuum and the weak $B_0$ correction,
can be written as
\begin{eqnarray} \label{GA-ff-B}
G_A^B (Q) &=& 
G_A^{M^*} (Q) \; +  \; 
\frac{eB_0}{{M^*}^2} 
\frac{F_{A}^{B} (0,Q)}{(eB_0/{M^*}^2)}. 
\\
 \label{gps-Bfield} 
G_{ps}^B (Q) &=& 
G_{ps}^{M^*} (Q) \; +  \; 
\left(\frac{eB_0}{{M^*}^2}\right)^2 
\frac{F_{ps}^{B} (0,Q)
}{(eB_0/{M^*}^2)^2}. 
\end{eqnarray}
Where $G_A^{M^*} (Q)$
 is the 
form factor presented and investigated 
 in \cite{FLB-2018a},  $G_A^{M^*}(Q)=G_A^U(0,Q)$,
 and similarly   $G_{ps}^{M^*} (Q) = G_{ps}(0,Q)$. 
In figure (\ref{fig:GPS-GA})
the axial and pseudoscalar 
non truncated
form factors 
in the vacuum - from \cite{FLB-2018a} - and 
with a weak magnetic field $eB_0/{M^*}^2 = 0.2$
from expressions (\ref{GA-ff-B},\ref{gps-Bfield})
are presented 
 for the two gluon propagators
with an unique value of the quark effective mass
$M^*=0.31$GeV.
Note however the $B_0$ correction to the pseudoscalar form factor 
has a factor 
$\left(\frac{eB_0}{{M^*}^2}\right)^2$ that is 
considerably smaller 
than 
$\left(\frac{eB_0}{{M^*}^2}\right)$
in the axial form factor.

\begin{figure}[ht!]
\centering
\includegraphics[width=140mm]{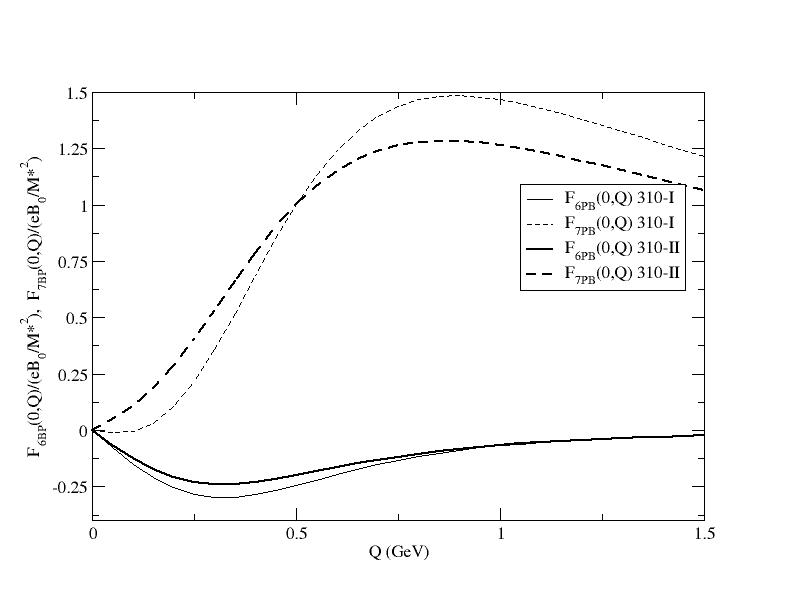}
%% {F6PB-F7PB-310-I+II.jpg}
%% F6B-CO-TM.jpg}
\caption{ \label{fig:F6PB-F7PB}
\small
The leading magnetic field  induced  
 form factors for the   pion coupling to   quark currents
$F_{6P-B} (0,Q), F_{7P-B} (0,Q)$, divided by $(eB_0)/{M^*}^2$,
as  functions  of the pion momentum, $Q=|Q_x|$,
are  presented 
for the two gluon   propagators $D_I (k)$  (thin lines) and 
 $D_{II}(k)$ (thick lines)
and  $M^*=310$MeV. 
 $h_I=0.82$ and $h_{II}=0.3$
}
\end{figure}

The unusual  weak magnetic field induced 
anisotropic 
 form factors, 
$F_{6PB}(0,Q)$ and $F_{7PB}(0,Q)$, are 
exhibitted in figure (\ref{fig:F6PB-F7PB})
as function of the pion momentum , $Q=|Q_x|$,
for the two gluon propagators with $M^*=310$MeV.
They disappear in the zero pion  momentum limit.
The dependence of the form factor $F_{7PB}(0,Q)$ on
the gluon propagator is seemingly larger than for the previous 
form factors analysed in the present work.
Although the order of magnitude
might be  larger than the corresponding
 electromagnetic form factors (\ref{LA-pij})
these values must be multiplied by 
$(eB_0)/{M^*}^2$.

%%%%%%%%%%%%%%%%%%%%%%%%%%%%
%%%%%%%%%%%%%%%%%%%%%%%%%%%%%%
%%%%%%%%%%%%%%%%%%%%%%%%%%%%%%%%%%%%%%

\subsection{ Averaged  quadratic  radii}

 From the above form factors,
 weak-$B_0$ induced  anisotropic corrections 
to the  axial and pseudoscalar 
constituent quark averaged  quadratic radii (a.q.r.)
can be
 obtained.
%  for expressions 
%% (\ref{q-pi-W-B},
%(\ref{q-pi-U-B}).
The magnetic field along the $\hat{z}$ direction
 can be chosen to be 
$A_\mu= -B_0 (0,y,x,0)/2$ 
for which it can be obtained a symmetrized result.
Because the $B_0$ corrections to the form factors 
are dimensionless as defined above, the a.q.r.
 can  be defined  simply as:
\begin{eqnarray} \label{Delta-rqm-B}
\Delta 
   < r^2 >_{A} &=& 
\left.
-  6 \; 
%\frac{6}{F_{\pi A-B}^U (0,0)}
 \frac{ d F_{A}^{B} (0,Q_\pi) }{d Q^2_\pi } \right|_{Q_\pi=0}
,
\nonumber
\\
\Delta 
 <  r^2  >_{ps}    &=& \left.
-  6  \; 
% \frac{6}{F_{psph-B}^U (0,0)}
 \frac{ d F_{ps}^{B} (0,Q_\pi) }{d Q^2_\pi } \right|_{Q_\pi=0}
.
\end{eqnarray}
that correspond to corrections  
 in the plane $x-y$ 
perpendicular to the constant weak magnetic field.
The corresponding a.q.r.  in the vacuum by considering the same
method have  been  presented in Ref. \cite{FLB-2018a}.
The resulting  value for the
axial and pseudoscalar square radii are obtained by 
adding their values in the  vacuum  to the magnetic field correction.
The
quark  effective mass $M^*$ however  is kept constant in
spite of its eventual  magnetic field dependence.
These values 
are obtained  by:
\begin{eqnarray}
 \label{rA-B}
< r^2 >_A^{B}   &=&  < r^2 >_A +
\left.
\left( \frac{eB_0}{{M^*}^2} \right) 
\frac{\Delta <r^2>_{A}}{(eB_0/{M^*}^2)} \right|_{x-y},
\\ 
 \label{rps-B}
< r^2 >_{ps}^B  &=&   < r^2 >_{ps} +
\left.
\left(\frac{eB_0}{{M^*}^2}\right)^2 
\frac{\Delta <r^2>_{ps} }{(eB_0/{M^*}^2)^2} \right|_{x-y},
\end{eqnarray}
where it was emphasized
 the magnetic field corrections stand only in  the plane perpendicular 
to the weak constant  magnetic field.
 There are  corrections to the
(strong) vector and scalar square radii
that  are similar and proportional to these
axial and pseudoscalar ones
 as obtained from the expressions 
%% (\ref{ff-W-vec-B},\ref{ff-W-sca-B},
(\ref{ff-sca-B},\ref{ff-U-vec-B}),
with slightly  different numerical factors.
They are related by:
\begin{eqnarray} \label{aqr-VA-SPS}
\left.
\Delta <r^2>_{V}  \right|_{x-y}
&=& 
3\left.
\Delta <r^2>_{A}   \right|_{x-y} ,
\nonumber
\\
\left.
\Delta <r^2>_{s}   \right|_{x-y}
&=& 
\frac{5}{12} \left.
\Delta <r^2>_{ps}   \right|_{x-y} .
\end{eqnarray}
  Therefore although the vector and axial form factors,
and the scalar and pseudoscalar form factors, in the vacuum
are equal, the corrresponding corrections due to
the weak $B_0$ are different as it could be expected 
because of the explicit isospin and  chiral  symmetry breakings 
due to $B_0$.

\begin{figure}[ht!]
\centering
\includegraphics[width=140mm]{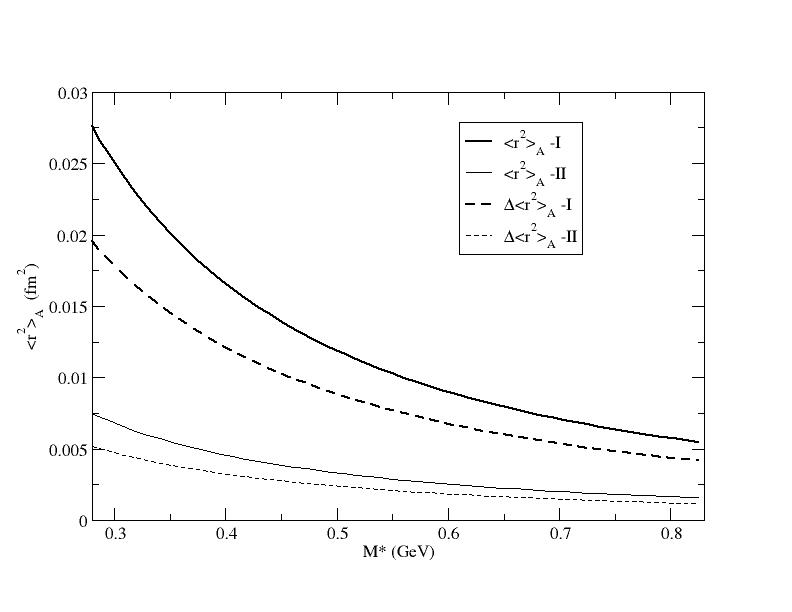}
% {rqm-gA-U+W-CO-fm2-ha.jpg}
% {rqm-gA-II-DB-nov.jpg}
%% {rqm-gA-U+W-CO-fm2.jpg}
\caption{ \label{fig:RQMA-B}
\small
The axial squared  radius from Ref. \cite{PRD-2018b} (thin and thick
 continuous 
lines)
 and anisotropic weak  magnetic field induced corrections
(dashed lines)
for  the  gluon propagators $D_I(k)$ and   $D_{II}(k)$ 
 as functions of the effective quark mass from the 
gap equation. $h_{II}=0.3$.
The weak magnetic field corrections from expressions
 (\ref{Delta-rqm-B}) are 
shown divided by the factor $eB/{M^*}^2$.
}
\end{figure}

In figure (\ref{fig:RQMA-B})
the axial quadratic  radii  extracted from \cite{FLB-2018a}
are compared  with the 
anisotropic $B_0$  induced contributions above (\ref{Delta-rqm-B})
as functions of the quark effective mass
for each of the gluon propagators.
The values of the  magnetic field  induced 
anisotropic contribution exhibitted 
 in these figures
must be multiplied by $eB_0/{M^*}^2$ 
to be added to the values $<r^2>_A$  in the vacuum 
as shown in expression (\ref{rA-B}).
  These axial and vector Strong quadratic radii 
yield smaller contributions  than those
found from the couplings to 
the light axial and vector mesons \cite{PRD-2018a},
and they are to be 
added.
 For the sake of comparison it is interesting to
quote previous estimations of the 
constituent quark quadratic radii to be 
$<r^2> \simeq 0.2-0.3$fm$^2$ \cite{weise-vogl,CQ-size},
being the relative  weak $B_0$ correction is of the relative  order of  
magnitude of
 $eB_0/{M^*}^2$
with respect to the corresponding value in the vacuum.

\begin{figure}[ht!]
\centering
\includegraphics[width=140mm]{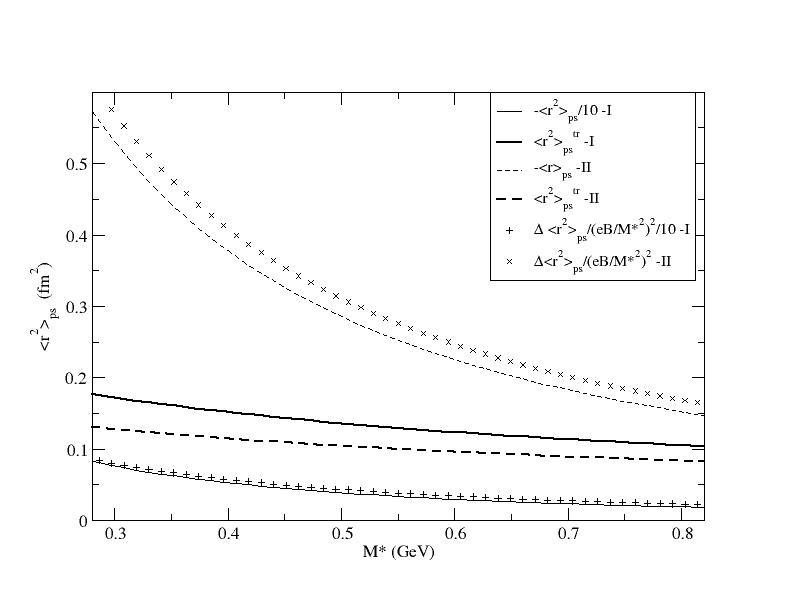}
%% {rqm-gps-I+II-nov.jpg}
%% {rqm-ps-I+II+tr-fm2.jpg}
\caption{ \label{fig:RQM-ps-B}
\small
The pseudoscalar squared  radius,
untruncated and truncated expressions,
from Ref. \cite{PRD-2018b} (continuous 
and dashed lines)
and the anisotropic magnetic field induced correction
 for the two 
gluon propagators, with $+$ and $\times$ ,
 as functions of the effective quark mass from the 
gap equation.
The case of untruncated $<r^2>_{ps}$ have a sign minus and
the case of $D_I(k)$ 
is divided by a factor 10.
 $h_I=0.82$ and $h_{II}=0.3$
The weak magnetic field corrections
are
shown divided by the factor $(eB/{M^*}^2)^2$.
}
\end{figure}

Finally the weak $B_0$ 
 pseudoscalar square radius, $<r^2>_{ps}$
from  Ref. \cite{FLB-2018a}  for the   truncated and non truncated
expressions,
 and the anisotropic $B_0$ induced correction,  
$\Delta <r^2>_{ps}$, from (\ref{Delta-rqm-B}),
 are exhibitted in figure (\ref{fig:RQM-ps-B})
as functions of the quark  gap effective mass $M^*$
 for $D_I(k)$ and $D_{II}(k)$.
It is noted however   that the anisotropic 
weak magnetic field corrections $\Delta_B <r^2>_{ps}$
yield very small values with respect to their values in the vacuum
 because of the factor $(eB_0/{M^*}^2)^2$.

\section{Summary}

The leading electromagnetic form factors of pion-constituent quark 
effective interactions 
and also
 corresponding relatively 
weak magnetic field corrections to constituent quark strong form factors
were derived within  a dynamical approach.
Besides the electromagnetic couplings of the 
 usual scalar, pseudoscalar, vector and axial 
pion effective interactions with constituent quarks,
two unusual photon couplings  were found,
$F_6$ and $F_7$.
All these effective couplings also provide 
  weak magnetic field anisotropic
corrections to the usual pion couplings,
besides 
 additional ones,  
inexistent in the vacuum, $F_{6B} (K,Q), F_{7B}(K,Q)$.
Two types of  weak magnetic field correction were found.
The usual case in which the  quark kernel receives a (linear)
anisotropic correction
from the weak magnetic field that is in the $\hat{z}$ direction
and 
the  case in which a  photon  coupled to 
the pion-constituent quark vertex  gives rise to
 a (relatively) weak magnetic field. 
The resummation of   higher
order  terms of the expansions done above
 can be expected to 
 provide the strong magnetic field regime
case
\cite{weak-B}. 
 The (relatively) weak magnetic field expansion however
allows for extracting analytical expressions for the 
related effective couplings that make explicit the 
involved physical effects behind observables.
Two anomalous Lorentz  pseudoscalar  pion couplings with vector
or axial constituent quark currents (\ref{S1-an})
% that are not Lorentz scalar but rather pseudoscalar terms
were  also found.
Two different non perturbative gluon propagators, which are known to 
produce DChSB, were shown to produce
specific smaller  differences in the behavior of  form factors with
pion momenta.
Because there are no current experimental measurements or 
theoretical estimations of 
 the magnetic field contribution to the baryons' form factors, in particular
 the axial and pseudoscalar ones, no further comparisons were possible
although these 
% it would be interesting to investigate these magnetic field
corrections may  have  some role  in 
the nucleon and in the  nuclear potentials  and in dense stars structure.
The pion contributions  to the axial, and vector, form factors
were found  however 
 to be smaller than the light axial and vector form contributions
\cite{PRD-2018a} and this conclusion holds for the magnetic field corrections.
The complete calculation for the nucleon form factors starting from 
the present dynamical approach was left outside the scope of the present work.
 In all these estimations the quark effective mass 
from the gap equation was kept 
constant, i.e. momentum independent,
 with the large contribution from the scalar condensate constant.
Form factors expressions were shown with the complete 
quark kernel momentum structure
  and in a truncated form. 
The main reason for truncating the expressions 
is that the resulting a.q.r. might be negative because of the
corresponding positive slope of  the 
complete expressions of form factors.
It was pointed out in \cite{FLB-2018a} this can be expected 
to be due to the absence of momentum dependence of 
the quark effective mass $M^*$ from the gap equation.
Finally,
weak magnetic field anisotropic  corrections to  averaged
 quadratic axial and pseudoscalar radii were also calculated
as functions of the quark effective mass.
The  vector and axial form factors corrections due to
the weak $B_0$ are not equal to each other 
 as it could be expected 
because of the explicit isospin and  chiral  symmetry breakings 
due to $B_0$.
 Up and down quark masses, however, were kept equal
in spite of the fact that their effective masses must be different 
and their couplings to the magnetic field are different.
This non degeneracy will  introduce corrections.
Finally axial and pseudoscalar 
averaged quadratic radii  were calculated as functions of the 
quark effective mass $M^*$.
These quadratic radii decrease
considerably  with the values of $M^*$  from $0.28$GeV to $0.78$GeV
although the different   gluon propagators, and the 
truncation of the form factors,  might yield
different slopes and normalizations.
The difficulties  of 
establishing  unbambiguous or precise values and  behavior for the 
quark gluon running coupling constant and 
non perturbative gluon propagator manifest mainly
 in the 
ambiguity of fixing the normalization
 values  for the zero momentum 
values  of the form factors.
However the relative values 
of the anisotropic corrections induced by $B_0$,
as compared to the corresponding quantities
in the vacuum, $F_A(K,Q)$,$F_{ps}(K,Q)$ 
and also  $<r^2>_A$,$<r^2>_{ps}$, 
are smaller basically 
by factors $eB_0/{M^*}^2$ or  $(eB_0/{M^*}^2)^2$
respectively.
The more general calculation for 
 strong magnetic fields is intended to be investigated elsewhere.

\section*{Acknowledgments}

F.L.B. thanks  short discussions 
with 
G.I. Krein,  P. Bedaque, I.Shovkhovy and G. Eichmann.
 F.L.B. participates to
 the project INCT-FNA, Proc. No. 464898/2014-5.

\end{document}